\documentclass[%
 reprint,
superscriptaddress,
 amsmath,amssymb,
pra,
]{revtex4-2}

\usepackage{amsmath}
\usepackage {xcolor}
\usepackage{comment}

\usepackage{mathrsfs}
\usepackage{comment}
\usepackage{amssymb}
\usepackage{graphicx}
\usepackage{subfigure}
\usepackage[colorlinks,
            linkcolor=blue,
            anchorcolor=blue,
            citecolor=blue]{hyperref}
\newtheorem{theorem}{Theorem}
\newtheorem{lemma}{Lemma}

\def\ra{\rangle}
\def\la{\langle}

\usepackage{graphicx}% Include figure files
\usepackage{dcolumn}% Align table columns on decimal point
\usepackage{bm}% bold math
\begin{document}

\preprint{APS/123-QED}

\title{Entropic uncertainty relations with quantum memory in a multipartite scenario}

\author{Qing-Hua Zhang}
\email[]{qhzhang@csust.edu.cn}

%\homepage[]{Your web page}
%\thanks{}
\affiliation{School of Mathematics and Statistics, Changsha University of Science and Technology, Changsha 410114, China}

\author{Shao-Ming Fei}
\email[]{feishm@cnu.edu.cn}
\affiliation{School of Mathematical Sciences, Capital Normal University,
Beijing 100048, China}
\affiliation{Max-Planck-Institute for Mathematics in the Sciences, 04103 Leipzig, Germany}

\begin{abstract}
Entropic uncertainty relations demonstrate the intrinsic uncertainty of nature from an information-theory perspective. Recently, a quantum-memory-assisted entropic uncertainty relation for multiple measurements was proposed by Wu $et\ al.$ [Phys Rev A. 106. 062219 (2022)]. Interestingly, the quantum-memory-assisted entropic uncertainty relation for multiple measurement settings can be further generalized. In this work, we propose two complementary multipartite quantum-memory-assisted entropic uncertainty relations and our lower bounds depend on values of complementarity of the observables, (conditional) von-Neumann entropies, Holevo quantities, and mutual information. As an illustration, we provide several typical cases to exhibit that our bounds are tighter and outperform the previous bounds.
\end{abstract}

\maketitle
%\tableofcontents

\section{Introduction}
The most revolutionary departure of quantum mechanics from classical mechanics is that it is impossible to simultaneously measure two complementary variables precisely. The original Heisenberg uncertainty relation was about the position and the momentum of a particle~\cite{heisenberg1927uber}. Robertson~\cite{robertson1929the} generalized the variance-based uncertainty relation for position and momentum to any two observables $M_1$ and $M_2$, $\Delta M_1  \Delta M_2 \geqslant \frac{1}{2} |\la \psi | [M_1, M_2] | \psi \ra |$, where $\Delta$ is the standard deviation of the observable with respect to a fixed state $|\psi\ra$ and $[M_1, M_2]$ represents the commutator of the observables $M_1$ and $M_2$.
Let $|\psi_j\ra$ and $|\phi_k\ra$ represent the eigenvectors of observables $M_1$ and $M_2$,  respectively. Deutsch~\cite{deutsch1983uncertainty} introduced an entropic uncertainty relation based on the Shannon entropy,
\begin{equation}
H(M_1)+H(M_2)\geqslant 2\log_2(\frac{2}{1+\sqrt{c}}),
\end{equation}
where $c=\max_{jk}|\langle \psi_j|\phi_k\rangle|^2$, $H(M_1)=-\sum_i p_i\log_2 p_i$ is the Shannon entropy with $p_i=\langle \psi_i|\rho|\psi_i\rangle$, and $H(M_2)=-\sum_i q_i\log_2 q_i$ with $q_i=\langle \phi_i|\rho|\phi_i\rangle$. Later, Kraus~\cite{kraus1987complementary}, Maassen, and Uffink~\cite{maassen1988generalized} improved Deutsch's result,
\begin{equation}\label{mu}
H(M_1)+H(M_2)\geqslant -\log_2 c=: q_{MU}.
\end{equation}

The famous quantum-memory-assisted entropic uncertainty relation (QMA-EUR) was introduced by Renes $et\ al.$~\cite{renes2009conjectured} and Berta $et\ al.$~\cite{berta2010uncertainty}, and verified by several well-designed experiments~\cite{li2011experimental,prevedel2011experimental}. The relation can be illustrated by so-called quantum game with two players Alice and Bob.
Initially, Bob prepares a two-particle state $\rho^{AB}$ and sends the particle $A$ to Alice. Then, Alice randomly chooses $M_1$ or $M_2$ to measure her part and announces her choice to Bob. If Bob guesses the measurement outcome correctly, he wins the game. The Bob's uncertainty in guessing Alice's measurement outcome is quantified by conditional von Neumann entropy. Mathematically, the minimum of the Bob's uncertainty is bounded by the following uncertainty relation,
\begin{equation}\label{renesberta}
S(M_1|B)+S(M_2|B)\geqslant -\log_2 c+S(A|B),
\end{equation}
where $ S(M|B)=S(\rho^{MB})-S(\rho^B)$ denotes the conditional von Neumann entropy of the postmeasurement state $\rho^{MB}=\sum_i(|\psi_i\ra\la\psi_i|\otimes \mathbb{I}) \rho^{AB} (|\psi_i\ra\la\psi_i|\otimes \mathbb{I})$, $S(A|B)=S(\rho^{AB})-S(\rho^B)$, $\rho^B$ is reduced state of particle $B$, and $S(\rho)=-{\rm Tr}\rho\log\rho$ is von Neumann entropy. The lower bound on the uncertainty of the measurement outcomes depends on the amount of entanglement between the measured particle $A$ and the quantum memory $B$. If the memory $B$ is absent, the above inequality becomes $ H(M_1)+H(M_2)\geqslant -\log_2 c+S(\rho^A)$, which yields a tighter lower bound comparing with (\ref{mu}) when $S(\rho^A)>0$. The QMA-EUR has many potential applications in various quantum information processing tasks, such as entanglement witness~\cite{berta2014entanglement,huang2010entanglement,hu2012quantum}, EPR steering~\cite{sun2018demonstration,schneeloch2013einstein}, quantum metrology~\cite{giovannetti2011advances}, quantum key distribution~\cite{koashi2009simple,berta2010uncertainty}, quantum cryptography~\cite{dupuis2014entanglement,konig2012unconditional}, and
quantum randomness~\cite{vallone2014quantum}.

Tighter lower bounds of QMA-EURs have been then investigated ~\cite{pati2012quantum,coles2014improved,adabi2016tightening}. In \cite{pati2012quantum} Pati $et\ al.$ verified that classical correlation and quantum correlation can strengthen the lower bound of QMA-EUR. In addition, Coles and Piani~\cite{coles2014improved} optimized the uncertainty relation with quantum memory by considering the second largest overlap of eigenvectors of the two observables measured. Adabi $et\ al.$~\cite{adabi2016tightening} optimized the lower bound of QMA-EUR by using the mutual information,
\begin{equation}\label{adabilb}
S(M_1 | B)+S(M_2 | B) \geqslant-\log _2 c+S(A | B)+\max \{0, \delta\},
\end{equation}
where $\delta=\mathcal{I}(A: B)-[\mathcal{I}(M_1: B)+\mathcal{I}(M_2: B)]$, $\mathcal{I}(A:$ $B)=S\left(\rho^A\right)+S\left(\rho^B\right)-S\left(\rho^{A B}\right)$ stands for the mutual information, $\mathcal{I}(M_1: B)=S\left(\rho^{M_1}\right)+S\left(\rho^B\right)-S\left(\rho^{M_1 B}\right)$ and $\mathcal{I}(M_2: B)=S\left(\rho^{M_2}\right)+S\left(\rho^B\right)-S\left(\rho^{M_2 B}\right)$ are the Holevo quantities.

Moreover, EUR for multiple measurements has attracted much attention after some excellent simulations based on two measurements were reported~\cite{zhang2015entropic,coles2017entropic,hu2013competition,ming2020improved,chen2018improved,xie2021optimized,liu2015entropic,xiao2016strong,dolatkhah2020tightening,dolatkhah2022tripartite}. Specially, Renes $et\ al.$~\cite{renes2009conjectured} and Berta $et\ al.$~\cite{berta2010uncertainty} established a tripartite entropic uncertainty relation with two memories $B$ and $C$,
\begin{equation}\label{reneslb}
S(M_1 | B)+S(M_2 | C) \geqslant q_{MU}.
\end{equation}
The uncertainty relation (\ref{reneslb}) describes a quantum game with three players: Alice, Bob, and Charlie. They agree with a common three-particle state and two measurements $M_1$ or $M_2$ on Alice's side. Alice randomly carries out one of the measurements $M_1$ or $M_2$ and announces her choice $M_1$ $(M_2)$ to Bob (Charlie). If both Bob and Charlie guess the measurement outcome correctly, they win the game. The uncertainty in guessing is quantified by conditional von Neumann entropy. The task of Bob and Charlie is to minimize the uncertainty.

Ming $et\ al.$~\cite{ming2020improved} presented an improved tripartite QMA-EUR by considering mutual information and the Holevo quantity,
\begin{equation}
S(M_1 | B)+S(M_2 | C) \geqslant q_{M U}+\max \{0, \delta_1\},
\end{equation}
where $\delta_1=2 S(A)+q_{M U}-\mathcal{I}(A: B)-\mathcal{I}(A: C)+\mathcal{I}(M_2: B)+$ $\mathcal{I}(M_1: C)-H(M_1)-H(M_2)$, which provides a tighter lower bound than that of Refs.~\cite{renes2009conjectured,berta2010uncertainty} and is of basic importance to enhance the security of quantum key distribution protocols. Later, Wu  $et\ al.$~\cite{wu2022tighter} improved the above bound further,
\begin{equation}
S(M_1 | B)+S(M_2 | C) \geqslant q_{M U}+\max \{0, \delta_2\},
\end{equation}
where $\delta_2=2 S(A)+q_{M U}-\mathcal{I}(M_1: B)- \mathcal{I}(M_2: C)-H(M_1)-H(M_2)$.

Quantum information processing may require to estimate the measurement uncertainty of not only two observables in bipartite or tripartite systems, but also multiple measurements in correlated multipartite systems. Recently,  Wu  $et\ al.$~\cite{wu2022tighter} also established a QMA-EUR,
\begin{equation}\label{wulb}
\sum_{i=1}^m S\left( {M}_i \mid B_i\right) \geqslant-\frac{1}{m-1} \log _2\left(\prod_{i\neq j}^{m} c_{ij}\right)+\max \left\{0, \delta_m\right\}
\end{equation}
with
$$
\begin{aligned}
\delta_m=& -\frac{1}{m-1} \log _2\left(\prod_{i\neq j}^{m} c_{ij}\right)+n S(A)-\sum_{i=1}^m H\left( {M}_i\right)\\&-\sum_{i=1}^m \mathcal{I}\left( {M}_i: B_i\right),
\end{aligned}
$$
where $ {\mathcal{M}}_i$ denotes the $i$ th measurement on subsystem $A$ and $B_i$ represents the $i$ th quantum memory in the multipartite system, and $c_{ij}=\max _{k,l}|\langle \psi_k^i |\psi_l^j\rangle|^2$ with $|\psi_k^i\rangle$ and $|\psi_l^j\rangle$ the eigenvectors of ${M}_i$ and $M_j$, respectively. Remarkably, it is found that the first item of the right-hand side of (\ref{wulb}) should be $-\frac{1}{m-1} \log _2\left(\prod_{i<j}^{m} c_{ij}\right)$, so does $\delta_m$.

More generally, motivated by the result of  Wu  $et\ al.$ in Ref.~\cite{wu2022tighter}, we consider entropic uncertainty relations in the context of $m$ measurements and $n$ memories with $n\leqslant m$. The lower bounds of these relations depend on the complementarity among the observables, the (conditional) von-Neumann entropies, the Holevo quantities and the mutual information. Furthermore, we derive a uniform QMA-EUR for multiple measurements within multipartite systems.

\section{Multipartite QMA-EURs for multiple measurements}

\begin{figure}[tp]
\includegraphics[width=7.8cm]{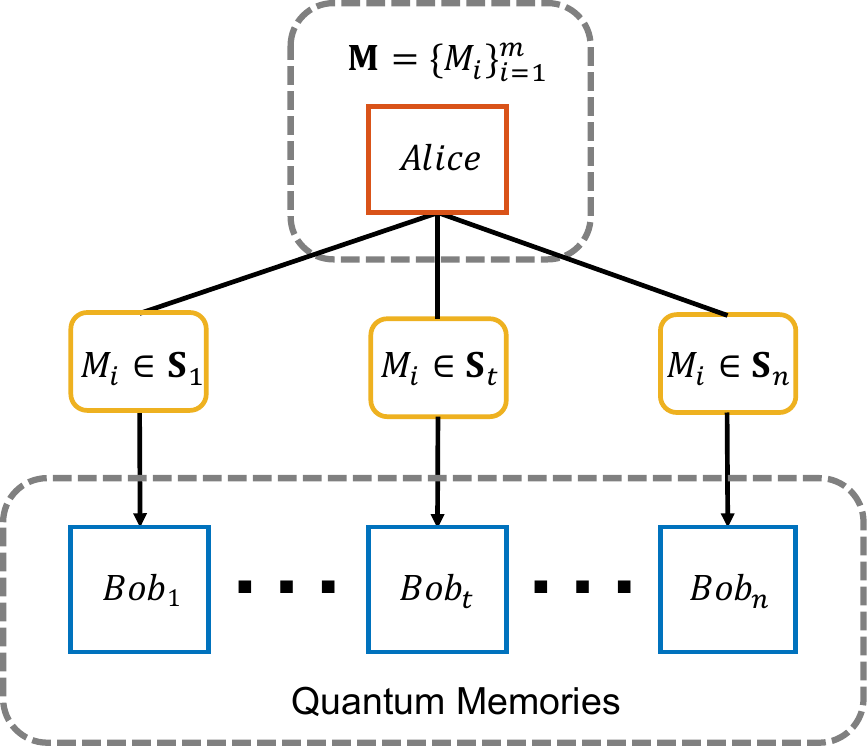}%
\caption{\label{qmaeur} Illustration of the uncertainty game in multipartite system. Alice carries out one measurement $M_i$ of $\mathbf{M}$ and announces her choice to Bob$_t$ if  $M_i\in \mathbf{S}_t$. The task of these Bob is to minimize the uncertainty about guessing Alice's measurement outcome.}
\end{figure}

Consider the following uncertainty game with $n+1$ players Alice, Bob$_1$, Bob$_2$, $\dots$, Bob$_n$: all players agree with a shared $n+1$ partite quantum state. Let $d$ be the dimension of Alice's partition. Alice randomly measures her part with one of $m-$tuple of measurements $\mathbf{M}={\{M_i\},\ i=1,\dots,m}$. Define $n$ non-empty subsets $\mathbf{S}_t$ of $\mathbf{M}$, such that $\bigcup_{t=1}^n \mathbf{S}_t=\mathbf{M}$ and $\mathbf{S}_s \bigcap \mathbf{S}_t=\emptyset$ for $s\neq t$. Alice carries out one measurement of  $\mathbf{S}_t$ and announces her choice to Bob$_t$. If all Bobs guess the measurement outcome correctly, they will win the game. The task of these Bobx is to minimize the uncertainty about guessing Alice's measurement outcome (see Fig.~\ref{qmaeur}).

We extend the entropic uncertainty relation (\ref{reneslb}) to more general uncertainty relations for the case of multiple measurements.
Similar to the additivity of linear uncertainty relations for local measurements~\cite{wehner2010entropic,schwonnek2018additivity,xie2021optimized}, we have the following simply constructed bound (SCB) of QMA-EUR for multiple measurements.

\begin{lemma}\label{eurlemma}
The generalized entropic uncertainty relation for $m$ measurements in the context of $n$ memories is given by
\begin{equation}\label{zhangeur1}
\begin{aligned}
\sum_{t=1}^n\sum_{M_i\in \mathbf{S}_t} S(M_i|B_t)\geqslant  &-\frac{1}{m-1} \log _2\left(\prod_{i<j}^{m} c_{ij}\right)\\
&+\frac{1}{m-1}
\sum_{t=1}^{n} \frac{m_t(m_t-1)}{2}S(A|B_t)\\
&=:U_{SCB},
\end{aligned}
\end{equation}
where $m_t$ is the cardinality of $\mathbf{S}_t$.
\end{lemma}
{\textit{Proof.}}
Rewriting the QMA-EURs (\ref{renesberta}) and (\ref{reneslb}), we have
\begin{equation*}
S(M_i|B_t)+S(M_j|B_s)\geqslant \left\{
\begin{aligned}
&-\log_2 c_{ij} +S(A|B_t),~ t=s\\
&-\log_2 c_{ij},~ t\neq s.
\end{aligned}
\right.
\end{equation*}
Summarizing these inequalities for all $i<j$ and dividing both sides by $m-1$, one proves the lemma. $\Box$

\begin{theorem}\label{qmaeurthm1}
Taking into account the conditional von-Neumann entropies, mutual information, and Holevo quantities, we have the following
entropic uncertainty relation for $m$ measurements in the context of $n$ memories:
\begin{equation}
\begin{aligned}
\sum_{t=1}^n\sum_{M_i\in \mathbf{S}_t} S(M_i|B_t)\geqslant &-\frac{1}{m-1} \log _2\left(\prod_{i<j}^{m} c_{ij}\right)\\
&+\frac{1}{m-1}\sum_t \frac{m_t(m_t-1)}{2}S(A|B_t)\\
&+\max\{0,\delta_{mn}\},
\end{aligned}
\end{equation}
where
$$
\begin{aligned}
\delta_{mn}=&\frac{m(m-1)-\sum_{t=1}^n m_t(m_t-1)}{2(m-1)}S(A)\\
&+\sum_{t=1}^n\frac{m_t(m_t-1)}{2(m-1)}\mathcal{I}(A:B_t)\\
&-\sum_{t=1}^n\sum_{M_i\in \mathbf{S}_t} \mathcal{I}(M_i:B_t).
\end{aligned}
$$
\end{theorem}

{\textit {Proof.}} For any two measurements $M_i$ and $M_j$ $(i\neq j)$, the uncertainty relation can be expressed by
\begin{align*}
&S(M_i|B_s)+S(M_j|B_t)\\
=&H(M_i)-\mathcal{I}(M_i:B_s)+H(M_j)-\mathcal{I}(M_j:B_t)\\
\geqslant& -\log_2(c_{ij}) +S(A)-\mathcal{I}(M_i:B_s)-\mathcal{I}(M_j:B_t),
\end{align*}
where the inequality is due to $H(M_i)+H(M_j)\geqslant -\log_2(c_{ij}) +S(A)$. One gets $m(m-1)/2$ analogous relations for all $i<j$. Summarizing these inequalities and dividing both sides by $m-1$, we obtain
\begin{equation}
\begin{aligned}
\sum_{t=1}^n\sum_{M_i\in \mathbf{S}_t} S(M_i|B_t)\geqslant &-\frac{1}{m-1}\log _2\left(\prod_{i<j}^{m} c_{ij}\right)\\
&+\frac{m}{2}S(A)-\sum_{t=1}^n\sum_{M_i\in \mathbf{S}_t} \mathcal{I}(M_i:B_t).
\end{aligned}
\end{equation}
The proof is completed by combining the above inequality with Lemma~\ref{eurlemma}. $\Box$

Specially, for a bipartite state $\rho^{AB}$ with $m$ measurements applied on particle $A$, that is, $n=1$, our uncertainty relation covers the QMA-EUR proposed by Xie $et\ al.$~\cite{xie2021optimized},
\begin{equation}\label{xie_lb}
\begin{aligned}
\sum_{i=1}^m S\left(M_i \mid B\right) \geqslant  &-\frac{1}{m-1} \log _2\left(\prod_{i<j}^{m} c_{ij}\right)+\frac{m}{2} S(A|B)\\
&  +\max \left\{0, \delta_{m1}\right\},
\end{aligned}
\end{equation}
where $\delta_{m1}=\frac{m}{2} \mathcal{I}(A: B)-\sum_{i=1}^m \mathcal{I}\left(M_i:B\right)$.

For the case of $m+1$ partite system with $m$ measurements applied to subsystem $A$ ($n=m$), the cardinality of $\mathbf{S}_t$ should be $1$, namely, $m_t=1$, our QMA-EUR in Theorem \ref{qmaeurthm1} induces to
\begin{equation}
\sum_{i}^m S(M_i|B_i)\geqslant -\frac{1}{m-1} \log _2\left(\prod_{i<j}^{m} c_{ij}\right)+\max\{0,\delta_{mm}\},
\end{equation}
where
$\delta_{mm}=\frac{m}{2}S(A)-\sum_{i}^m \mathcal{I}(M_i:B_i)$.

In Ref.~\cite{dolatkhah2022tripartite}, Dolatkhah $et\ al.$ proposed  tighter tripartite QMA-EURs in terms of the EURs for multiple measurements. Motivated by this, we generalize the method to multipartite systems.

\begin{theorem}\label{qmaeurthm2}
The following entropic uncertainty relation for $m$ measurements in the context of $n$ memories holds,
\begin{equation}
\begin{aligned}
\sum_{t=1}^n\sum_{M_i\in \mathbf{S}_t} S(M_i|B_t)\geqslant &-\frac{1}{m-1} \log _2\left(\prod_{i<j}^{m} c_{ij}\right)\\
&+\frac{1}{m-1}\sum_t \frac{m_t(m_t-1)}{2}S(A|B_t)\\
&+\max\{0,\delta_{mn}^\prime\},
\end{aligned}
\end{equation}
where
\begin{equation*}
\begin{aligned}
\delta_{mn}^\prime=&\frac{1}{m-1}\log_2\frac{\left(\prod_{i<j}^{m} c_{ij}\right)}{b^{m-1}}+(m-1)S(A)\\
&-\sum_{t=1}^n\frac{m_t(m_t-1)}{2(m-1)}S(A)+\sum_{t=1}^n\frac{m_t(m_t-1)}{2(m-1)}\mathcal{I}(A:B_t)\\
&-\sum_{t=1}^n\sum_{M_i\in \mathbf{S}_t} \mathcal{I}(M_i:B_t).
\end{aligned}
\end{equation*}
\end{theorem}

{\textit{Proof.}}
From the conditional von-Neumann entropy $S(M_i|B_t)=H(M_i)-\mathcal{I}(M_i:B_t)$, we have
\begin{equation}
\begin{aligned}
\sum_{i=1}^m H(M_i)=\sum_{t=1}^n\sum_{M_i\in \mathbf{S}_t} S(M_i|B_t)+\sum_{t=1}^n\sum_{M_i\in \mathbf{S}_t} \mathcal{I}(M_i:B_t).
\end{aligned}
\end{equation}
In Ref.~\cite{liu2015entropic}, Liu $et\ al.$ proposed an entropic uncertainty relation for multiple measurements based on quantum channels,
\begin{equation}\label{liulb}
\sum_{i=1}^m H\left(M_i\right) \geqslant-\log_2 b+(m-1) S(A),
\end{equation}
where
$$
b=\max _{k_m}\left\{\sum_{k_2 \sim k_{m-1}} \max _{k_1} |\langle \psi_{k_1}^1 | \psi_{k_2}^2\rangle |^2 \prod_{i=2}^{m-1} |\langle \psi_{k_i}^i| \psi_{k_{i+1}}^{i+1}\rangle |^2 \right\}.
$$
with $\sum_{k_2 \sim k_{m-1}}=\sum_{k_2}\sum_{k_3}\cdots\sum_{k_{m-1}}$. Using (\ref{liulb}), one reaches
\begin{equation}
\begin{aligned}
\sum_{t=1}^n\sum_{M_i\in \mathbf{S}_t} S(M_i|B_t) \geqslant&-\log
_2 b+(m-1) S(A)\\
&-\sum_{t=1}^n\sum_{M_i\in \mathbf{S}_t} \mathcal{I}(M_i:B_t).
\end{aligned}
\end{equation}
Combining with (\ref{zhangeur1}), we complete the proof. $\Box$

In particular, for the case of bipartite system with $m$ measurements applied to subsystem $A$ ($n=1$), our uncertainty relation becomes
\begin{equation}
\begin{aligned}
\sum_i^mS(M_i|B)\geqslant &-\frac{1}{m-1} \log _2\left(\prod_{i<j}^{m} c_{ij}\right)+\frac{m}{2}S(A|B)\\
&+\max\{0,\delta_{m1}^\prime\},
\end{aligned}
\end{equation}
where $\delta_{m1}^\prime=\frac{1}{m-1}\log_2\frac{\left(\prod_{i<j}^{m} c_{ij}\right)}{b^{m-1}}+\frac{m-2}{2}S(A)
+\frac{m}{2}\mathcal{I}(A:B)-\sum_i^m \mathcal{I}(M_i:B)$.
It is interesting to compare the lower bounds of Theorems \ref{qmaeurthm1} and \ref{qmaeurthm2} in this scenario. Set
\begin{equation}
\delta_{m1}^\prime-\delta_{m1}=\frac{1}{m-1}\log_2\frac{\left(\prod_{i<j}^{m} c_{ij}\right)}{b^{m-1}}+\frac{m-2}{2}S(A).
\end{equation}

Mutually unbiased observables are of importance in quantum information theory \cite{nielsen2002quantum}. The observables $M_1$ and $M_2$ are called mutually unbiased if their eigenbases $\{|\psi^{M_1}_i\rangle\}_{i=1}^d$ and $\{|\psi^{M_2}_j\rangle\}_{j=1}^d$ are mutually unbiased, namely, $|\langle \psi^{M_1}_i|\psi_j^{M_2}\rangle|^2=1/d$, $\forall\ i,j$. When the measurements are taken to be the mutually unbiased ones, one has $c_{ij}=b=1/d$. Thus we have
\begin{equation}
\delta_{m1}^\prime-\delta_{m1}=\frac{2-m}{2}\log_2d+\frac{m-2}{2}S(A)\leqslant 0,
\end{equation}
where the inequality is due to $S(A)\leq \log_2d$. when one takes mutually unbiased observables, the lower bound of Theorem \ref{qmaeurthm1} is equivalent to that of Theorem \ref{qmaeurthm2} for $m=2$ and is strictly tighter than that of Theorem \ref{qmaeurthm2} for $m\geqslant 3$.

For the case of $m+1$ partite system with $m$ measurements applied to subsystem $A$ ($n=m$), the cardinality of $\mathbf{S}_t$ should be $1$, that is to say, $m_t=1$, our QMA-EUR reduces to
\begin{equation}
\sum_{i}^m S(M_i|B_i)\geqslant -\frac{1}{m-1} \log _2\left(\prod_{i<j}^{m} c_{ij}\right)+\max\{0,\delta_{mm}\},
\end{equation}
where $\delta_{mm}=\frac{1}{m-1}\log_2\frac{\left(\prod_{i<j}^{m} c_{ij}\right)}{b^{m-1}}+(m-1)S(A)-\sum_{i}^m \mathcal{I}(M_i:B_i)$.
Similarly, when one employs mutually unbiased observables,
\begin{equation}
\delta_{mm}^\prime-\delta_{mm}=\frac{2-m}{2}\log_2d+\frac{m-2}{2}S(A)\leqslant 0,
\end{equation}
where the inequality is due to $S(A)\leq \log_2d$. The lower bound of Theorem \ref{qmaeurthm1} is also tighter than that of Theorem \ref{qmaeurthm2} for mutually unbiased observables.

Generally, we have the following uniformly constructed QMA-EUR for multiple measurements in the context of multipartite system.

\begin{theorem}\label{qmaeurthm3}
Let $\sum_i^mH(M_i)\geqslant U$ be a general form of Shannon entropic uncertainty relations for $m$ measurements $\mathbf{M}=\{M_i\}_{i=1}^m$. Then the following multipartite QMA-EUR for multiple measurements holds,
\begin{equation}
\begin{aligned}
\sum_{t=1}^n\sum_{M_i\in \mathbf{S}_t} S(M_i|B_t)\geqslant &U_{SCB}+\max\{0,U-U_{SCB}\\
&-\sum_{t=1}^n\sum_{M_i\in \mathbf{S}_t} \mathcal{I}(M_i:B_t)\}.
\end{aligned}
\end{equation}
\end{theorem}

The proof is similar to that of Theorem \ref{qmaeurthm2}. Following Theorem \ref{qmaeurthm3}, the tighter $U$ means tighter lower bound of the constructed multipartite QMA-EUR.

\section{Performance of our QMA-EURs}
In this section, we illustrate the performance of our QMA-EURs by three typical cases. Since the measured particle in the following is a qubit, the mutually unbiased Pauli matrices $\sigma_x,\ \sigma_y$ and $\sigma_z$ can be chosen as observables.

\subsection{QMA-EURs for one memory}

Alice and Bob agree on Pauli matrices $\sigma_x,\ \sigma_y$ and $\sigma_z$ as measurements. Alice carries out one of these measurements and announces her choice to Bob. Bob's task is to minimize the uncertainty about the outcome of Alice's measurement.
In this case, our Theorem~\ref{qmaeurthm1} reduces to
\begin{equation}
\begin{aligned}
&S\left(\sigma_x | B\right)+S\left(\sigma_y | B\right)+S\left(\sigma_z | B\right)\\
 \geqslant & \frac{3}{2}+\frac{3}{2} S(A|B)
+\max \left\{0, \frac{3}{2} \mathcal{I}(A: B)-\!\!\!\sum_{i=x,y,z}\!\! \mathcal{I}\left(\sigma_i:B\right)\right\},
\end{aligned}
\end{equation}
which is coincident with the QMA-EUR proposed by Xie $et\ al.$~\cite{xie2021optimized}. Our Theorem~\ref{qmaeurthm2} reduces to
 \begin{equation}
\begin{aligned}
&S\left(\sigma_x | B\right)+S\left(\sigma_y | B\right)+S\left(\sigma_z | B\right)\\
\geqslant  &\frac{3}{2}+\frac{3}{2} S(A|B) +\max \left\{0, \delta_{31}^\prime\right\},
\end{aligned}
\end{equation}
where $\delta_{31}^\prime=-\frac{1}{2}+\frac{1}{2}S(A)+\frac{3}{2}\mathcal{I}(A\!:\!B)-\sum_{i=x,y,z} \mathcal{I}\left(\sigma_i\!:\!B\right)$.

\begin{figure}[tbp]
	\begin{subfigure}
	 \centering
	\label{figex1a}
		\includegraphics[width=4.2cm]{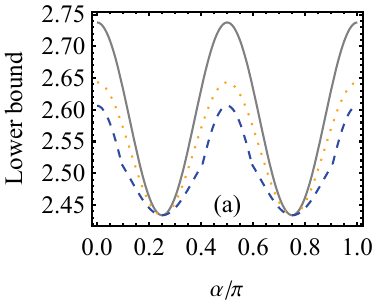}
% Add this line to your code
	\end{subfigure}
	\begin{subfigure}
		\centering
		\label{figex1b}
		\includegraphics[width=4cm]{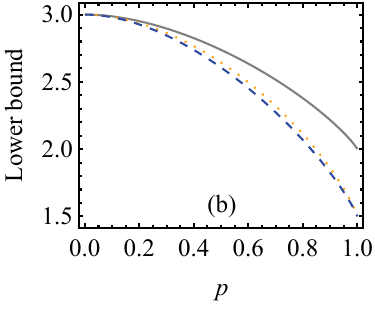}
% Add this line to your code
	\end{subfigure}
% \subfigure[]
% {
% \label{figex1b} %% label for second subfigure
% \includegraphics[]{ex1b.pdf}}
 \caption{Uncertainty and the lower bounds vs $\alpha$ and $p$ for two-qubit mixed state $\rho=p|\sigma\ra\la \sigma|+(1-p)\frac{\mathbb{I}_{4}}{4}$,
where $|\sigma\ra=\cos{\alpha}|00\ra+\sin{\alpha|11\ra}$. The gray (solid) curve represents the uncertainty. The yellow (dotted) and blue (dashed) curves denote the lower bounds of Theorem~\ref{qmaeurthm1} and~\ref{qmaeurthm2}, respectively.}
 \label{figex1}
 \end{figure}

As an example, let us consider a class of two-qubit mixed states, $\rho=p|\sigma\ra\la \sigma|+(1-p)\frac{\mathbb{I}_{4}}{4}$,
where $0\leqslant p\leqslant 1$, $|\sigma\ra=\cos{\alpha}|00\ra+\sin{\alpha|11\ra}$ with $\alpha \in [0,2\pi)$, and $\mathbb{I}_{4}$ denotes the $4\times 4$ identity matrix. When we take into account the Pauli matrices $\sigma_x,\ \sigma_y$ and $\sigma_z$ as observables. As shown in Fig.~\ref{figex1}{\color{blue}(a)} for $p=\frac{1}{2}$ and in Fig.~\ref{figex1}{\color{blue}(b)} for $\alpha=\frac{\pi}{2}$, the lower bound of our Theorem~\ref{qmaeurthm1} is tighter than that of Theorem~\ref{qmaeurthm2}, which coincides with the uncertainty relation in Ref.~\cite{xie2021optimized}, tighter than that of Ref.~\cite{liu2015entropic} for mutually unbiased observables.

\subsection{QMA-EURs for two memories}

Alice, Bob and Charlie agree on Pauli matrices $\sigma_x,\sigma_y$ and $\sigma_z$ as observables. If Alice carries out $\sigma_x$, Bob guesses the outcome of Alice's measurement. If Alice carries out $\sigma_y$ or $\sigma_z$, then Charlie guesses the result of Alice's measurement.
In this case, our Theorem~\ref{qmaeurthm1} becomes
\begin{equation}
\begin{aligned}
&S\left(\sigma_x | B\right)+S\left(\sigma_y | C\right)+S\left(\sigma_z | C\right)\\
\geqslant&   \frac{3}{2}+\frac{1}{2}S(A|C)+\max \left\{0, \delta_{32}\right\},
\end{aligned}
\end{equation}
where $\delta_{32}=S(A)-\frac{1}{2}\mathcal{I}(A:C)-\mathcal{I}\left(\sigma_x:B\right)-\mathcal{I}\left(\sigma_y:C\right)-\mathcal{I}\left(\sigma_z:C\right)$. Our Theorem~\ref{qmaeurthm2} reduces to
\begin{equation}
\begin{aligned}
&S\left(\sigma_x | B\right)+S\left(\sigma_y | C\right)+S\left(\sigma_z | C\right)\\
&\geqslant  \frac{3}{2}+\frac{1}{2}S(A|C)+\max \left\{0, \delta_{32}^\prime\right\},
\end{aligned}
\end{equation}
where $\delta_{32}^\prime=\frac{-1}{2}+\frac{3}{2}S(A)+\frac{1}{2}\mathcal{I}(A:C)-\mathcal{I}\left(\sigma_x:B\right)-\mathcal{I}\left(\sigma_y:C\right)-\mathcal{I}\left(\sigma_z:C\right)$.

\begin{figure}[tbp]
\centering
	\begin{subfigure}
	 \centering
	\label{figex1a}
		\includegraphics[width=4cm]{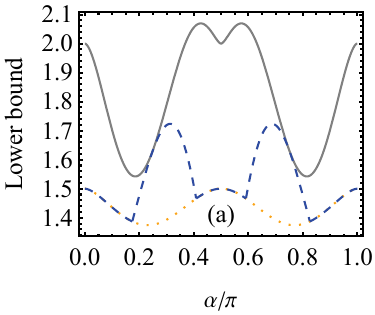}
% Add this line to your code
	\end{subfigure}
	\begin{subfigure}
	 \centering
	\label{figex1a}
		\includegraphics[width=4cm]{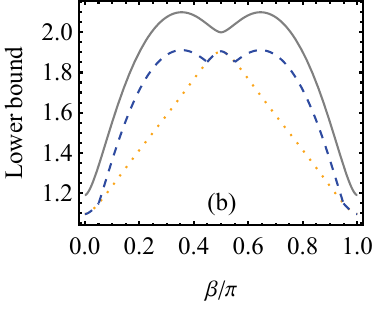}
% Add this line to your code
	\end{subfigure}
\caption{Uncertainty and the lower bounds vs $\alpha$ and $\beta$ for the generalized W state. The gray (solid) curve represents the uncertainty. The yellow (dotted) curve and
blue (dashed) curve denote the lower bounds of Theorems~\ref{qmaeurthm1} and~\ref{qmaeurthm2}, respectively.}
\label{figex2}
\end{figure}

Let us consider the generalized W state,
$|W\ra=\sin{\alpha}\cos{\beta}|001\ra+\sin{\alpha}\sin{\beta}|010\ra+\cos{\alpha}|100\ra$
where $\alpha\in [0,\pi)$ and $\beta\in [0,2\pi)$. The comparison between our Theorem~\ref{qmaeurthm1} and \ref{qmaeurthm2} is shown in Fig.~\ref{figex2}. Figure~\ref{figex2}{\color{blue}(a)} shows the case of $\beta=\frac{\pi}{5}$ and Fig.~\ref{figex2}{\color{blue}(b)} shows the case of $\alpha=\frac{2\pi}{3}$. In this example the lower bound of Theorem~\ref{qmaeurthm2} is strictly tighter than that of Theorem~\ref{qmaeurthm1}.

\subsection{QMA-EURs for three memories}

Alice, Bob, Charlie and David agree on measurements $\sigma_x,\ \sigma_y$ and $\sigma_z$. Alice carries out one of these measurements and announces her choice to others.
Bob, Charlie and David guess results of measurement $\sigma_x$, $\sigma_y$ and $\sigma_z$, respectively.
In this case, our Theorem~\ref{qmaeurthm1} reduces to
\begin{equation}
\begin{aligned}
S\left(\sigma_x | B\right)+S\left(\sigma_y | C\right)+S\left(\sigma_z | D\right)
\geqslant  \frac{3}{2}+\max \left\{0, \delta_{31}\right\}
\end{aligned}
\end{equation}
where $\delta_{31}=\frac{3}{2} S(A)-\mathcal{I}\left(\sigma_x:B\right)-\mathcal{I}\left(\sigma_y:C\right)-\mathcal{I}\left(\sigma_z:D\right)$ and Theorem~\ref{qmaeurthm2} gives rise to
\begin{equation}
\begin{aligned}
S\left(\sigma_x | B\right)+S\left(\sigma_y | C\right)+S\left(\sigma_z | D\right) \geqslant  \frac{3}{2}+\max \left\{0, \delta_{31}^\prime\right\},
\end{aligned}
\end{equation}
where $\delta_{31}^\prime=-\frac{1}{2}+2S(A)-\mathcal{I}\left(\sigma_x:B\right)-\mathcal{I}\left(\sigma_y:C\right)-\mathcal{I}\left(\sigma_z:D\right)$.

To illustrate the performance of our theorems, we compare with the results in Ref.\cite{wu2022tighter}. The uncertainty relation (\ref{wulb}) of Wu $et\ al.$ can be rewritten as
\begin{equation}
\begin{aligned}
S\left(\sigma_x | B\right)+S\left(\sigma_y | C\right)+S\left(\sigma_z | D\right) \geqslant  \frac{3}{2}+\max \left\{0, \delta_{3}\right\},
\end{aligned}
\end{equation}
where $\delta_{3}=\frac{3}{2}+3S(A)-H(\sigma_x)-H(\sigma_y)-H(\sigma_z)-\mathcal{I}\left(\sigma_x:B\right)-\mathcal{I}\left(\sigma_y:C\right)-\mathcal{I}\left(\sigma_z:D\right)$.

\begin{figure}[tbp]
 \centering
 \subfigure[]
 {
 \label{figex3a}
		\includegraphics[width=4.1cm]{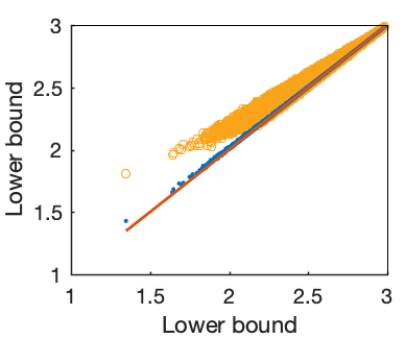}}
 \subfigure[]
 {
 \label{figex3b}
 \includegraphics[width=4.1cm]{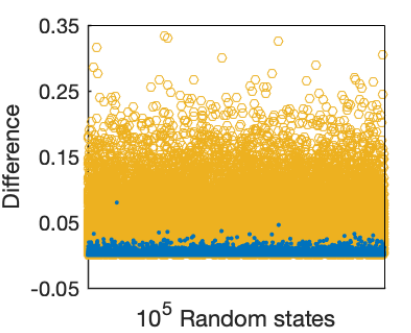}}
 \caption{(a) Comparison between our bounds and the bound in Ref.~\cite{wu2022tighter} for $10^5$ random states. The $y$ axis is lower bound. The bound in Ref.~\cite{wu2022tighter} is just the $x$ axis. The yellow circles and blue dots denote the lower bounds of Theorem~\ref{qmaeurthm1} and Theorem~\ref{qmaeurthm2}, respectively. (b) The $x$ axis is for random states and the $y$ axis is the difference between our lower bounds and the bound in Ref.~\cite{wu2022tighter}. The yellow circles denote the lower bound of Theorem~\ref{qmaeurthm1} minus the bound in Ref.~\cite{wu2022tighter}. The blue dots represent the lower bound of Theorem~\ref{qmaeurthm2} minus the bound in Ref.~\cite{wu2022tighter}.}
 \label{figex3}
 \end{figure}

To be more general, we consider arbitrary sets of random four-qubit states for this case.~According to the spectral decomposition theorem, an arbitrary four-qubit state can be decomposed into $\rho^{ABCD}=\sum_{k=1}^{16}p_k|\psi_k\ra\la\psi_k|$, where $p_k$ and $|\psi_k\ra$ denote the $i$th eigenvalue and eigenvector, respectively. The sets of normalized eigenvectors can be used to construct a unitary operation. An arbitrary four-qubit state can be attained by generating sets of probabilities and unitary operations. The random number function $f(0,1)$ gives independent random numbers generated uniformly in a closed interval $[0,1]$. On the one hand, one can effectively generate $16$ random probabilities $p_i$,
\begin{equation}
p_k=\frac{q_k}{\sum_{k=1}^{16}q_k},~ k=1,\dots,16,
\end{equation}
where $q_1=f(0,1)$ and $q_{k+1}=f(0,1)q_k$. In this way one obtains a set of probabilities in descent order.
On the other hand, one can randomly generate a $16$-order real matrix $R$ by using the random function $f(-1,1)$ within a closed interval $[-1,1]$. Based on the real matrix $R$, one can construct a random Hermitian matrix, $\tilde{R}=D+(U^\mathbb{T}+U)+i(L^\mathbb{T}+L)$, where $D$, $U$ and $L$, respectively, denote the diagonal, strictly upper and lower triangular parts of the real matrix $\tilde{R}$, $U^\mathbb{T}$ is the transpose of $U$. In this way one gets $16$ normalized eigenvectors of the random matrix $\tilde{R}$. Consequently, the spectral decomposition of a random four-qubit state can be perfectly constructed.
To verify our conclusions, we adopt $10^5$ random states to show our lower bounds and that of Wu's bound in Ref.~\cite{wu2022tighter}, as shown in Fig.~\ref{figex3}. From Fig.~\ref{figex3} we see that the lower bound of Theorem \ref{qmaeurthm1} is optimal.

\section{Conclusion}
We have proposed two complementary multipartite quantum-memory-assisted entropic uncertainty relations for multiple measurements. Furthermore, we have presented a uniform method to construct QMA-EURs according to the EURs for multiple measurements. As an illustration, we have considered the QMA-EURs for three measurements with one, two and three memories.

Although the QMA-EURs for two measurements can be applied directly for the security proof of quantum key distribution with two conjugate observables \cite{berta2010uncertainty}, our QMA-EURs are potentially to be applied to qualify the amount of secure keys in multipartite quantum key distribution. Experimentally, uncertainty inequalities without memory can be demonstrated by conventional quantum systems such as trapped ions~\cite{zhang2013state,duan2010colloquium} and photonic qudits~\cite{lapkiewicz2011experimental}. With the presence of quantum memory, the experimental realization can also be implemented with more than two measurements settings~\cite{li2011experimental,prevedel2011experimental}. Therefore, our proposals can be also experimentally implemented.

\bigskip
\noindent{\bf Acknowledgments}\, \,
This work is supported by the National Natural Science Foundation of China (NSFC) under Grants 12075159 and 12171044, Beijing Natural Science Foundation (Grant No. Z190005), Academician Innovation Platform of Hainan Province, and Changsha University of Science and Technology (Grant No. 000303923).

\nocite{*}
\bibliographystyle{apsrev4-2}
\bibliography{qmaeur}

\end{document}